\documentclass[twocolumn,aps,prc,superscriptaddress,showpacs,showkeys,floatfix,nofootinbib]{revtex4}
\usepackage{url}
\usepackage{cancel}
\usepackage[colorlinks,linkcolor=blue,citecolor=blue,filecolor=blue,urlcolor=blue]{hyperref}
\usepackage{epsfig,graphics}
\usepackage{graphicx}
\usepackage{dcolumn}
\usepackage{bm}
\usepackage[usenames]{color}
\usepackage{amssymb}
\usepackage{amsmath}
\usepackage{multirow}
\usepackage{float}
\usepackage{harpoon}
\usepackage{MnSymbol}
\usepackage{appendix}
\usepackage{color}
\usepackage{hyperref}
\usepackage{cleveref}
\usepackage{booktabs}  
\usepackage{orcidlink}
\usepackage{cleveref}
\usepackage{CJKutf8}

\newcommand{\sqrtsnn}{\mbox{$\sqrt{s^{}_{\mathrm{NN}}}$}}

\newcommand\be{\begin{equation}}
\newcommand\ee{\end{equation}}

\begin{document}
\title{Probing the neutron-skin thickness through $J/\psi$ photoproduction in ultra-peripheral collisions}

\author{Haoyuan Li}
\affiliation{Shanghai Institute of Applied Physics, Chinese Academy of Sciences, Shanghai $201800$, China}
\affiliation{University of Chinese Academy of Sciences, Beijing $101408$, China}
\affiliation{Key Laboratory of Nuclear Physics and Ion-beam Application (MOE), Fudan University, Shanghai $200433$, China}

\author{Lu-Meng Liu \orcidlink{0000-0001-5243-5549}}\email[Correspond to\ ]{liulumeng@fudan.edu.cn}
\affiliation{Physics Department and Center for Particle Physics and Field Theory, Fudan University, Shanghai $200438$, China}

\author{Jinhui Chen \orcidlink{0000-0001-7032-771X}}\email[Correspond to\ ]{chenjinhui@fudan.edu.cn}
\affiliation{Key Laboratory of Nuclear Physics and Ion-beam Application (MOE), Fudan University, Shanghai $200433$, China}
\affiliation{Shanghai Research Center for Theoretical Nuclear Physics, National Natural Science Foundation of China and Fudan University, Shanghai $200438$, China}

\author{Yu-Gang Ma \orcidlink{0000-0002-0233-9900}}\email[Correspond to\ ]{mayuganga@fudan.edu.cn}
\affiliation{Key Laboratory of Nuclear Physics and Ion-beam Application (MOE), Fudan University, Shanghai $200433$, China}
\affiliation{Shanghai Research Center for Theoretical Nuclear Physics, National Natural Science Foundation of China and Fudan University, Shanghai $200438$, China}
\affiliation{School of Physics, East China Normal University, Shanghai $200241$, China}

\author{Chunjian Zhang \orcidlink{0000-0002-5425-7130}}\email[Correspond to\ ]{chunjianzhang@fudan.edu.cn}
\affiliation{Key Laboratory of Nuclear Physics and Ion-beam Application (MOE), Fudan University, Shanghai $200433$, China}
\affiliation{Shanghai Research Center for Theoretical Nuclear Physics, National Natural Science Foundation of China and Fudan University, Shanghai $200438$, China}

\begin{abstract}
We study the impact of neutron-skin thickness on $J/\psi$ photoproduction in ultra-peripheral $^{208}\mathrm{Pb}+{}^{208}\mathrm{Pb}$ collisions. Within the Color Glass Condensate framework, we calculate coherent and incoherent cross sections and examine their dependence on the momentum transfer $|t|$ for different neutron-skin thicknesses. We find a clear imprint of the neutron skin on the $|t|$ spectra: a larger neutron skin leads to a smoother and more extended color-density profile, suppressing the coherent cross section at large $|t|$ while enhancing the incoherent cross section through increased event-by-event configurational fluctuations in the nuclear periphery.  We further show that the ratio of incoherent to coherent integrated cross sections provides a particularly sensitive and robust observable, with reduced theoretical uncertainties. These results establish diffractive vector-meson photoproduction in ultra-peripheral collisions as a powerful tomographic tool to constrain the neutron-skin thickness and the transverse gluon distribution at the LHC and future Electron–Ion Colliders.
\end{abstract}
\keywords{relativistic heavy-ion collisions; ultra-peripheral collisions; color glass condensate; neutron skin}
\maketitle

\section{Introduction} 
The spatial distributions of neutrons and protons in heavy nuclei are central to nuclear physics, offering profound insights into nuclear surface properties and nuclear astrophysics~\cite{Steiner:2004fi,Lattimer:2006xb,Li:2008gp}. In neutron-rich heavy nuclei, excess neutrons are pushed toward the periphery, forming a neutron skin, whose thickness \(\Delta r_{np}\) is defined as the difference between the root-mean-square (RMS) radii of the neutron and proton densities. The neutron-skin thickness serves as a sensitive probe of the slope of the nuclear symmetry energy and thus provides key constraints on the isovector sector of the nuclear equation of state (EoS)~\cite{Zhang:2017ncy,Chen:2010qx,An_NST,Hu:2021trw}. Traditional low-energy experimental measurements provide essential constraints on \(\Delta r_{np}\), such as parity-violating electron scattering~\cite{PREX:2021umo,Kumar:2020ejz}.

Recent heavy-ion collisions offer a unique and complementary experimental avenue to study this structure~\cite{Ma:2022dbh,STAR:2024wgy,STAR:2025elk,Zhang:2021kxj,Jia:2021oyt,Chen:2024aom,Chen:2026gka,Xi_NST,Giacalone:2024bud,SB_NST}. In such collisions, the pronounced neutron skin---manifested as an extended diffusion of neutrons beyond the proton distribution-modifies the initial nuclear geometry. This geometric modification, in turn, influences the subsequent collective dynamics and final-state observables in collisions at intermediate energies~\cite{Fang:2010zzc,Li:2016gjs,Ding:2024xxu,MaCW_NST,Yan_NST,Li:2024fsx}, high energies~\cite{Jia:2022qgl,Giacalone:2023cet,Li:2019kkh,Liu:2023pav}, and through spectator-based probes~\cite{Liu:2022xlm,Liu:2022kvz,Liu:2023qeq}. Consequently, quantifying the impact of the neutron skin on the spatial gluon distribution is essential for achieving a precision understanding of Quantum Chromodynamics (QCD) in the high-energy regime and for consistently extracting EoS parameters from experimental data~\cite{Jia:2022ozr,Sorensen:2023zkk}.

Ultra-peripheral collisions (UPCs) provide an alternative and particularly clean setting to probe initial-state nuclear structure through photonuclear interactions~\cite{STAR:2021wwq,STAR:2022wfe,Abelev_2014,CMS:2025oxg,ATLAS:2025aav,ATLAS:2025nac,ALICE:2024whv,CMS:2023pzm,ALICE:2023jgu,Brandenburg:2025one,Shen_Res,Jia:2025oak}. In UPCs, hadronic interactions are strongly suppressed, and the electromagnetic fields of relativistic ions can be described as a flux of quasi-real photons within the Equivalent Photon Approximation (EPA)~\cite{Baur2008,Baur:1998ay,starlight_Klein_2017}. In this regime, exclusive vector-meson production provides a tomographic probe of the transverse nuclear geometry and is sensitive to low-energy nuclear structure, including neutron-skin effects and nuclear deformation~\cite{Klein:2019qfb,Klein:2019avl,Lomnitz:2018juf,chun23,mantysaari2023,Zhao:2022dac,Wang:2024xeo}. This UPC program complements the future Electron-Ion Collider (EIC)~\cite{Accardi:2012qut,Anderle:2021wcy,Chu:2024fpo,Wu_NST,AbdulKhalek:2021gbh,Zhang:2025raf}, which will extend such studies to smaller Bjorken-$x$ and into the nonlinear regime of QCD~\cite{Accardi:2012qut,Boer:2011fh}. At small \(x\), the nucleus is probed as a gluon-dominated system, where the initial geometry and its fluctuations play an important role in scattering observables~\cite{chun23,mantysaari2023}.

At high energies, exclusive vector-meson production in UPCs can be described within the color-dipole framework~\cite{Gimeno-Estivill:2024gbu,Mantysaari:2022sux,Bartels:2003yj,Watt:2008vq}. In this approach, a quasi-real photon emitted by the projectile fluctuates into a quark-antiquark (\(q\bar{q}\)) dipole, which then scatters off the target nucleus. The squared momentum transfer is \(t=-|\boldsymbol{\Delta}_\perp|^2\), where \(\boldsymbol{\Delta}_\perp\) is Fourier conjugate to the center-of-mass coordinates of a dipole pair. The \(|t|\) dependence of the differential cross section \(\mathrm{d}\sigma/\mathrm{d}t\) therefore provides access to the transverse spatial structure of the target~\cite{Bertulani:2005ru,chun23}. Scattering proceeds through two channels: the coherent channel, where the nucleus remains in its ground state and the cross section probes the average nuclear density profile, and the incoherent channel, where the nucleus dissociates and the cross section is sensitive to event-by-event fluctuations of the nuclear configuration.

In this work, we study coherent and incoherent cross sections for diffractive \(J/\psi\) production in \(\gamma+\mathrm{Pb}\) collisions within the dipole picture, with an emphasis on quantifying their sensitivity to the neutron-skin thickness. The spatial modification induced by the neutron skin is reflected in the \(|t|\) dependence of the cross sections, making exclusive vector-meson production a useful tool for constraining the nuclear periphery. 

The remainder of this paper is organized as follows. Section~\ref{sec:theory} introduces the theoretical framework of the present study. Section~\ref{sec:results} presents and discusses numerical results for \(^{208}\mathrm{Pb}+{}^{208}\mathrm{Pb}\) collisions. Finally, a summary is given in Sec.~\ref{sec:summary}.

\section{Theoretical Framework}
\label{sec:theory}

The differential cross section for $J/\psi$ photoproduction as a function of the squared momentum transfer $|t|$ in $\gamma^*+\mathrm{Pb}\rightarrow J/\psi+\mathrm{Pb}^{(*)}$ collisions provides a high-resolution probe of nuclear structure at small Bjorken-$x$. In this work, we follow the framework developed in Refs.~\cite{chun23,mantysaari2023} to investigate the impact of the neutron skin. Below, we briefly outline the formalism for $J/\psi$ photoproduction in photon--nucleus interactions and in UPCs without interference effect, as well as the parametrization of the neutron skin.

\subsection{$J/\psi$ photon-nuclear production}
In scattering, the coherent cross section where the target nucleus remains in its initial state is proportional to the event averaged amplitude squared and normalized by \cite{PhysRev.120.1857,Klein:2023zlf}
\begin{equation} 
\frac{\mathrm{d}\sigma^{\gamma^* + \text{Pb} \rightarrow J/\psi + \text{Pb}}}{\mathrm{d}t} = \frac{1}{16\pi} \left| \langle \mathcal{A} \rangle_{\Omega} \right|^2 ,
\label{eq.coherent}
\end{equation}
where $\mathcal{A}$ denotes the photon nuclear scattering amplitude and $\langle \dots \rangle_{\Omega}$ represents the average over the color configurations $\Omega$ of the target.

The incoherent cross section corresponds to processes in which the target nucleus dissociates. It is defined as the difference between the total exclusive cross section and the coherent contribution~\cite{Mantysaari:2020axf,PhysRevD.18.1696,PhysRev.120.1857,Klein:2023zlf},
\begin{equation}
    \frac{\mathrm{d}\sigma^{\gamma^* + \text{Pb} \rightarrow J/\psi + \text{Pb}^*}}{\mathrm{d}t} = \frac{1}{16\pi}\left[\langle | \mathcal{A} |^2\rangle_\Omega - |\langle \mathcal{A} \rangle_\Omega|^2\right].
    \label{eq.incoherent}
\end{equation}
Physically, the incoherent cross section measures the variance of the scattering amplitude and thus provides direct access to event-by-event fluctuations of the nuclear density profile.

Within the Color Glass Condensate (CGC) framework~\cite{Iancu_2004,Gelis_2010,McLerran_2011}, which describes QCD dynamics in the small-$x$ regime, the exclusive vector-meson production amplitude is computed in the color-dipole picture~\cite{chun23,Ji:1996nm,Hatta:2017cte,Kowalski_2006,mantysaari2023}. The leading-order expression for the amplitude reads
\begin{equation} 
\begin{aligned} 
\mathcal{A} = 2i \int & \mathrm{d}^2\mathbf{s}_\perp\mathrm{d}^2\boldsymbol{r}_\perp\frac{\mathrm{d}z}{4\pi} \mathrm{e}^{-i\left[\mathbf{s}_\perp - \left(\frac{1}{2} - z\right)\boldsymbol{r}_\perp\right] \cdot \boldsymbol{\Delta}_\perp} \\ & \times [\psi^*_V \psi_\gamma]\left(Q^2, \boldsymbol{r}_\perp, z\right) \mathcal{N}_\Omega \left(\mathbf{s}_\perp,\boldsymbol{r}_\perp, z, x_\mathbb{P}\right).
\label{eq.amp}
\end{aligned} 
\end{equation}
Here \( [\psi_V^*\psi_\gamma](Q^2,\boldsymbol{r}_\perp,z) \) denotes the overlap between the incoming photon and outgoing vector‑meson wave functions, with \(Q^2\) the photon virtuality. In this work we consider quasi‑real photons and set \(Q^2=0\). The variables \(\boldsymbol{r}_\perp\) and \(z\) denote, respectively, the transverse dipole size and the longitudinal momentum fraction carried by the quark, while \(\mathbf{s}_\perp\) is the transverse distance from the target center to the dipole center. The quantity \(\mathcal{N}_\Omega\) is the dipole–nucleus scattering amplitude for a given color configuration \(\Omega\).

\begin{figure}[t]
\centering
\vspace*{-0.2cm}
\includegraphics[width=0.85\linewidth]{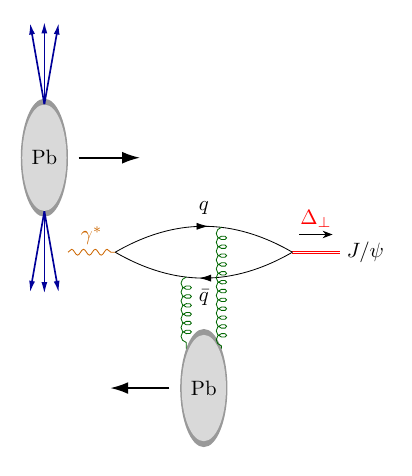}
\vspace*{-0.8cm}
\caption{Schematic diagram of the $J/\psi$ photon-nuclear production in ultra-peripheral $^{208}\text{Pb} + ^{208}\text{Pb}$ collision.}
\label{fig:fig1}
\end{figure}

The phenomenological factorization of the scattering process is illustrated in Fig.~\ref{fig:fig1}. A quasi-real photon, described via the EPA~\cite{Baur:1998ay,starlight_Klein_2017}, fluctuates into a color dipole $\gamma \rightarrow q\bar{q}$ that scatters off the target nucleus before hadronizing into the $J/\psi$ meson. The squared momentum transfer is given by $t=-\boldsymbol{\Delta}_\perp^2$. For the wave function overlap $\left[\psi_V^*\psi_\gamma\right]\left(Q^2, \boldsymbol{r}_\perp, z\right)$, we employ the boosted Gaussian parametrization with parameters tuned to decay data~\cite{Watt:2008vq,Kovchegov22}.

The dipole scattering amplitude can be expressed in terms of the dipole $\mathcal{S}$-matrix as~\cite{Hatta:2017cte}
\begin{equation}
\mathcal{N}_\Omega(\mathbf{s}_\perp,\boldsymbol{r}_\perp,z,x_\mathbb{P})
= 1 - \mathcal{S}_\Omega(\mathbf{s}_\perp,\boldsymbol{r}_\perp,z,x_\mathbb{P}),
\label{eq.dipoleTmatrix}
\end{equation}
where the $\mathcal{S}$-matrix is given by the correlator of Wilson lines.

In CGC formalism, Wilson line $V(x)=\mathcal{P}\operatorname{exp}\left[ -i\int \mathrm{d}x \mathcal{A}\left(x\right)\right]$ describes the interaction between quarks and the target nuclear color fields \cite{Iancu_2004}. The dipole $\mathcal{S}$-matrix can be written as
\begin{equation}
    \mathcal{S}_\Omega \left(\mathbf{s}_\perp ,\boldsymbol{r}_\perp , z,x_\mathbb{P}\right) = \frac{1}{N_C} \operatorname{Tr} \left[ V\left(\mathbf{s}_\perp +z\boldsymbol{r}_\perp\right) V^\dagger \left(\mathbf{s}_\perp - \left(1-z\right)\boldsymbol{r}_\perp\right) \right],
    \label{eq.dipoleSmatrix}
\end{equation}
where $\boldsymbol{r}_\perp$ denotes the transverse size of the $q\bar q$ dipole and $\mathbf{s}_\perp$ is the impact parameter of the photon relative to the center of the target nucleus. The corresponding dipole scattering amplitude is therefore
\begin{equation}
    \mathcal{N}\left(\mathbf{s}_\perp, \boldsymbol{r}_\perp,z, x_{\mathbb{P}}\right) = 1 - \frac{1}{N_C} \operatorname{Tr} \left[ V\left(\mathbf{s}_\perp + z\boldsymbol{r}_\perp\right) V^\dagger \left(\mathbf{s}_\perp -\left(1-z\right)\boldsymbol{r}_\perp\right) \right].
    \label{eq.dipoleamp}
\end{equation}
Within the CGC effective theory, the Wilson line is defined as
\begin{equation}
\begin{aligned}
    \mathrm{V}\left(\boldsymbol{x}_\perp\right) =\mathcal{P}_{-}\operatorname{exp} \left[ -ig\int \mathrm{d}x^-\frac{\rho^a \left(x^-, \boldsymbol{x}_\perp\right)t^a} {\boldsymbol{\operatorname{\nabla}}_\perp^2 - m^2}\right],
    \label{eq.Wilsonline}
\end{aligned}
\end{equation}
where $\mathcal{P}_-$ represents the path order along light cone time direction. In CGC EFT, the degree of freedom of color fields is understood as soft gluons in a background color source generated by hard gluons. The gauge fields are solutions to the Classical Yang-Mills equations with the color source density $\rho^a\left(x^-,\mathbf{x}_\perp\right)$ and $t^a$ are the $SU\left(3\right)_C$ gauge group generators. The $m^2$ term in the denominator serves as an infrared (IR) regulator to prevent infrared divergences.

In the context of the Wilson-line formalism and the factorization of the scattering amplitude in Eqs.~\eqref{eq.amp}, \eqref{eq.dipoleamp}, and \eqref{eq.Wilsonline}, the scattering amplitude probes the spatial distribution of gluons. In the dilute limit (i.e., the weak-field regime), the dipole scattering amplitude can be expanded in color density $\rho^a$ and its lowest non-trivial order $\rho^2$ corresponds to the two-gluon exchange approximation. Consequently, the coherent cross section is sensitive to the average color density of the target nucleus, while the incoherent cross section reflects event-by-event statistical fluctuations of the color density.

To account for sub-nucleonic fluctuations, we employ an event-by-event sampling approach for the initial color charge density. In this framework, the density profile of an individual nucleon is modeled as a collection of $N_q$ constituents, commonly referred to as ``hot spots"~\cite{Mantysaari:2017dwh}. The resulting nuclear thickness function, $T_p\left(\mathbf{s}_\perp\right)$, is defined as the weighted average of these constituents:
\begin{equation}
    T_p\left(\mathbf{s}_\perp\right) = \frac{1}{N_q} \sum_{i=1}^{N_q} p_i T_q\left(\mathbf{s}_\perp -\mathbf{s}_{\perp,i}\right).
    \label{Tp}
\end{equation}
Here, The coefficients $p_i$ serve as normalization weights for each individual hot spot. $T_q\left(\mathbf{s}_\perp\right)$ represents the transverse density profile of a single hot spot, characterized by a Gaussian distribution with width $B_q$:
\begin{equation}
    T_q(\mathbf{s}_\perp) = \frac{1}{2\pi B_q} e^{-{\mathbf{s}_\perp^2}/{2B_q}}.
    \label{Tq}
\end{equation}
In our event-by-event simulations, the spatial coordinates $\mathbf{s}_{\perp,i}$ of the $N_q$ hot spots are sampled from a broader Gaussian distribution with a characteristic width $B_{qc}$. In this work, the proton fluctuation parameters obtained are $B_q = 0.3\;\text{GeV}^{-2},\;B_{qc}=4.3\;\text{GeV}^{-2}$.
\subsection{$J/\psi$ production in UPC}
In ultra-peripheral heavy-ion collisions, the production of vector mesons can be factorized into two main processes: the emission of quasi-real photons, described by the Equivalent Photon Approximation, and the subsequent photon-nucleus interaction, which we evaluate within the Color Glass Condensate framework. According to the EPA, the equivalent photon flux per unit area and photon energy $\omega$ is given by~\cite{starlight_Klein_2017,Bertulani:2005ru}
\begin{equation}
    N(\omega, \mathbf{b}_\perp) = \frac{Z^2\alpha\omega^2}{\pi^2 \gamma^2}\left[K^2_1\left(\frac{\omega |\mathbf{b}_\perp|}{\gamma}\right) + \frac{1}{\gamma^2}K_0^2\left(\frac{\omega |\mathbf{b}_\perp|}{\gamma}\right)\right],
    \label{photon}
\end{equation}
where $\mathbf{b}_\perp$ is the transverse distance from the center of the emitting nucleus to the target (the nucleus-nucleus impact parameter), $Z$ is the nuclear charge number, $\gamma$ is the Lorentz boost factor, $\alpha \approx 1/137$ is the electromagnetic fine-structure constant, and $K_{0,1}$ are the modified Bessel functions of the second kind. 

In this work, we neglect the interference effects between the two photon-emitting nuclei and assume that the photons carry strictly longitudinal momentum (i.e., neglecting the transverse photon momentum). Under these simplifications, the UPC cross section factorizes into the independent product of the photon flux and the photon-nucleus cross section. To ensure the collision remains ultra-peripheral and free of hadronic interactions, the effective photon flux $n(\omega)$ is obtained by integrating $N(\omega, \mathbf{b}_\perp)$ over the transverse plane with the condition $|\mathbf{b}_\perp| > 2R_{\text{Pb}}$ (where $R_{\text{Pb}}$ is the nuclear radius)
\begin{equation}
    n(\omega) = \int_{|\mathbf{b}_\perp| > 2R_{\text{Pb}}} \mathrm{d}^2\mathbf{b}_\perp \, N(\omega, \mathbf{b}_\perp).
    \label{flux}
\end{equation}
Consequently, the differential cross section for vector meson ($J/\psi$) production in A-A collisions can be expressed as
\begin{equation}
    \frac{\mathrm{d}\sigma^{\text{PbPb} \rightarrow \text{Pb}\text{Pb}^{(*)} J/\psi}}{\mathrm{d}t} = n(\omega_1)\frac{\mathrm{d}\sigma^{\gamma\text{Pb}}}{\mathrm{d}t}(\omega_1) + n(\omega_2)\frac{\mathrm{d}\sigma^{\gamma\text{Pb}}}{\mathrm{d}t}(\omega_2).
    \label{totalcross}
\end{equation}
Here, the photon energies from the two colliding nuclei are \(\omega_{1,2}=\frac{M_V}{2}e^{\pm y}\), where \(y\) is the rapidity and \(M_V = 3.096\) GeV is the \(J/\psi\) mass. The corresponding gluon momentum fractions are \(x=\frac{M_V}{\sqrt{s_{\mathrm{NN}}}}e^{\mp y}\), and the photon–nucleon center-of-mass energy is \(W_{\gamma N}=\sqrt{M_V \sqrt{s_{\mathrm{NN}}}\,e^{\pm y}}\). In our calculation, we consider \(J/\psi\) production at \(\sqrt{s_{\mathrm{NN}}}=5.02\,\text{TeV}\) with \(y=-2,-1,0,1,2\).

It should be noted that in our calculations, there are some phenomenological uncertainties in the vector meson wave function and missing higher-order corrections in the dipole scattering amplitude~\cite{Mantysaari:2025ltq}. Meanwhile, the parameters we use in the IP-Glasma model are not fitted to experimental data. Therefore, in our calculations, the absolute cross-section values may not perfectly match the experimental data.

\begin{table}[htb]
    \centering
    \caption{Neutron skin parameter setting of $^{208}\text{Pb}$.}
    \label{tab:WSpara}
    \renewcommand{\arraystretch}{1.3}
    \setlength{\tabcolsep}{4mm}
    \begin{tabular}{ccccc}
        \hline\hline
        Case & $\delta R_{np}\;(\rm fm)$ & $\delta a_{np}$ & $\Delta r_{np}\;(\rm fm)$\\
        \hline
        $\text{\rm Case 1}$ & 0.01 & -0.3 & -0.225 \\
        $\text{\rm Case 2}$ & 0.0 & 0.0 & 0.0 \\
        $\text{\rm Case 3}$ & 0.01 & 0.155 & 0.210 \\
        $\text{\rm Case 4}$ & 0.01 & 0.3 & 0.445 \\
        $\text{\rm Case 5}$ & 0.60 & 0.0 & 0.445\\
        \hline\hline
    \end{tabular}
\end{table}

\subsection{Neutron skin setting}
To quantify the sensitivity of scattering observables to neutron skin effect in $^{208}\text{Pb}$, we model the spatial distribution of nucleons using a Woods-Saxon (WS) profile
\begin{equation}
\rho_\tau\left(r\right) = \frac{\rho_{\tau0}}{1+\exp\left[\left(r-R_\tau\right)/a_\tau\right]},
\label{eq.ws}
\end{equation}
where $\tau=n,p$ labels neutrons and protons, respectively. The parameters $R_\tau$ and $a_\tau$ control the nuclear radius and surface diffuseness, while $\rho_{\tau0}$ ensures proper normalization. For protons, we fix $R_p=6.68~\mathrm{fm}$ and $a_p=0.468$ (corresponding to
a nuclear rms radius $\sqrt{\langle r_p^2 \rangle} = 5.436\;\text{fm}$)~\cite{Giacalone:2023cet}. Possible nuclear‑deformation effects are neglected to isolate the impact of the neutron skin; a combined analysis of UPC and central‑collision observables (e.g., anisotropic flow) may be required to disentangle deformation effects from neutron‑skin effects.

In this work, five distinct scenarios are considered to systematically investigate the sensitivity of observables to the neutron skin by varying \(\delta R_{np}=R_n-R_p\) and the diffuseness difference \(\delta a_{np}=a_n-a_p\). Case~1 represents a hypothetical scenario where the neutron skin effect is suppressed (\(\Delta r_{np}<0\;\mathrm{fm}\)) for comparison; Case~2 serves as the baseline with no neutron skin (\(\Delta r_{np}=0.0\;\mathrm{fm}\)); and Case~3 is chosen to be consistent with current low‑energy experimental constraints (\(\Delta r_{np}=0.21\;\mathrm{fm}\))~\cite{PREX:2021umo}. Case~4 corresponds to an extreme limit of the neutron‑skin thickness (\(\Delta r_{np}=0.445\;\mathrm{fm}\)). Notably, Case~5 is introduced to incorporate the bulk component of the neutron skin~\cite{Warda:2010qa} by varying only the radius \(R_n\), allowing us to verify whether our observables are sensitive to bulk contributions. The parameter sets are listed in Table~\ref{tab:WSpara}.

Event-by-event Wilson lines for these density profiles are generated with the IP-Glasma framework~\cite{ipglasma_code,subnucleondiffraction_code,Schenke:2012wb,Schenke:2012hg}, using parameters from Ref.~\cite{chun23}. For each neutron‑skin scenario, we simulate \(N=5000\) independent configurations to compute \(d\sigma/dt\). For the integrated ratio \(\sigma_{\text{Incoh}}/\sigma_{\text{Coh}}\) and \(d\sigma/dy\), we use \(N=2000\) configurations. Statistical uncertainties are estimated with the bootstrap method, which suppresses finite‑sample bias and yields robust fluctuation estimates.

\section{Results and discussions} \label{sec:results}

\begin{figure}[t]
\centering
\includegraphics[width=0.9\linewidth]{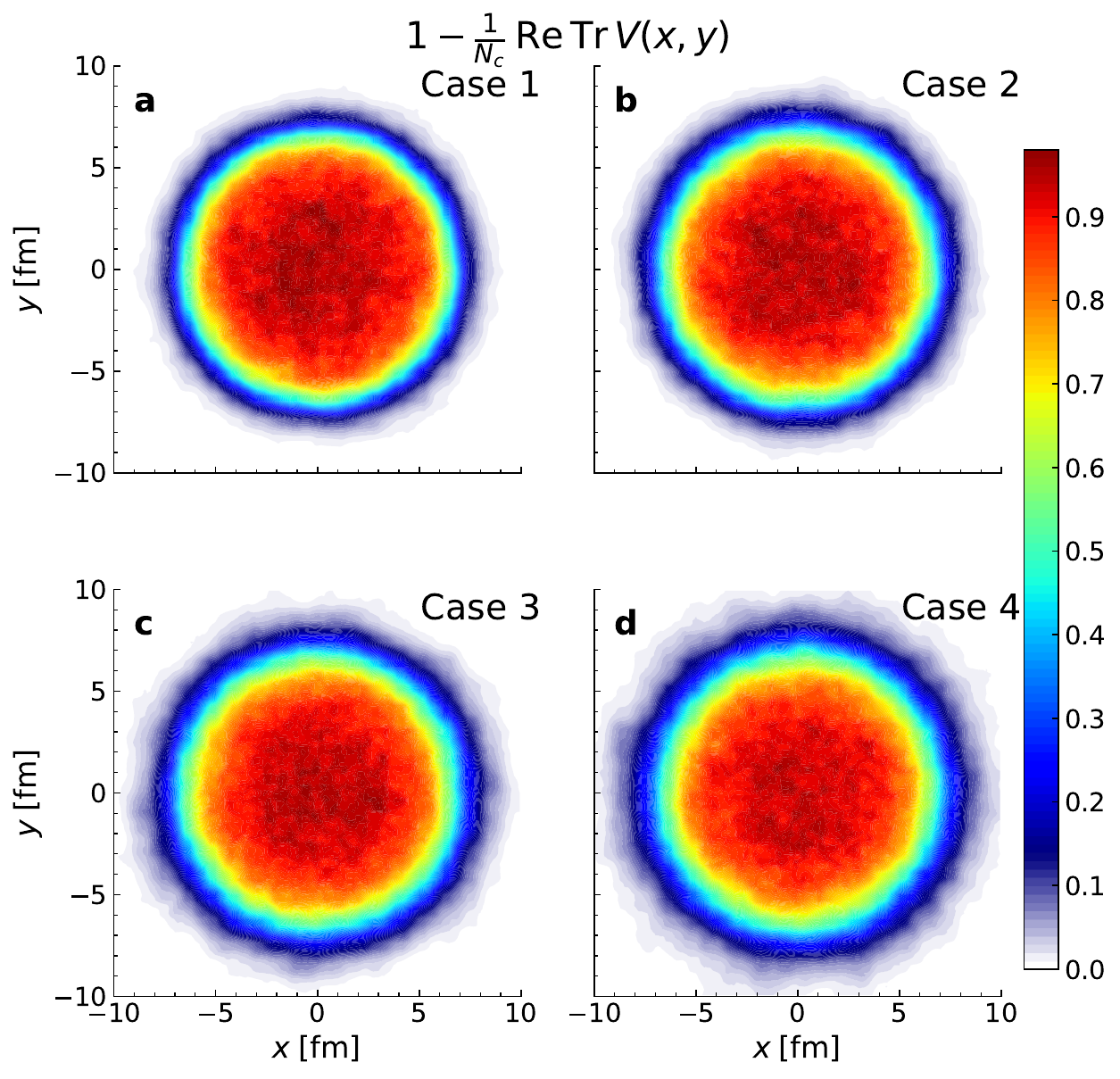}
\vspace{-0.3cm}
\caption{The nuclear color-density profile $1-\frac{1}{N_C}\operatorname{ReTr}V(x,y)$ in the CGC framework at $x\sim 0.001$ for different neutron-skin scenarios.}
\label{fig:fig2}
\end{figure}

Within the CGC framework, the local nuclear color density is described by the Wilson-line correlator \(1-\frac{1}{N_c}\operatorname{Re}\operatorname{Tr}V(x,y)\)~\cite{chun23,Gimeno-Estivill:2024gbu,Mantysaari:2022sux,Bartels:2003yj,McLerran_2011,Gelis_2010,Iancu_2004}. Figure~\ref{fig:fig2} illustrates these spatial profiles for the scenarios defined in Table~\ref{tab:WSpara}. As the neutron skin increases, the nuclear edge appears more diffuse. Such a geometric extension in the color distribution provides a natural structural basis for understanding the subsequent scattering observables.

\begin{figure}[htb]
\centering
\includegraphics[width=1\linewidth]{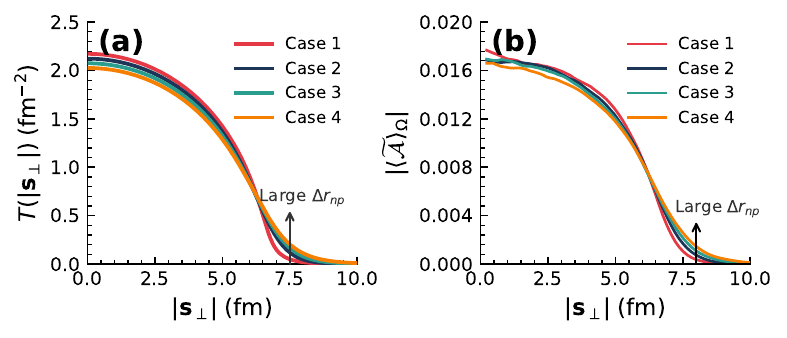}
\vspace{-0.7cm}
\caption{Nuclear thickness function $T(|\mathbf{s}_\perp|)$ of ${}^{208}\mathrm{Pb}$ (a) and the distribution $|\langle \widetilde{\mathcal{A}} \rangle_{\Omega}|$ in $\gamma+{}^{208}\mathrm{Pb}$ collisions at $x\sim 0.001$ (b), shown as functions of the transverse distance $|\mathbf{s}_\perp|$ for different neutron-skin scenarios.}
\label{fig:thickness}
\end{figure}

We evaluate the nuclear thickness function \(T(|\mathbf{s}_\perp|)=\int_{-\infty}^{\infty}\mathrm{d}z\,\rho(\sqrt{\mathbf{s}_\perp^2+z^2})\) (the nuclear density in the transverse plane) to characterize the transverse structure. As shown in Fig.~\ref{fig:thickness} (a), a larger neutron-skin thickness corresponds to a broader neutron distribution toward the periphery, featuring a slightly more diffuse central core and an extended tail at large transverse radius.
To further quantify this color density profile, we examine the scattering amplitude as a function of the impact parameter \(\mathbf{s}_\perp\), which is given by the Fourier transform of the scattering amplitude defined in Eq.~\eqref{eq.amp} and show the $\gamma+\text{Pb}$ scattering form factor distribution in transverse plane. It is defined as 
\[
\widetilde{\mathcal{A}}(\mathbf{s}_\perp)
= 2i \int \frac{\mathrm{d}^2\boldsymbol{r}_\perp\, \mathrm{d}z}{4\pi}
\left[\Psi_\gamma \Psi_V^*\right](\boldsymbol{r}_\perp, z)\,
\mathcal{N}_\Omega(\mathbf{s}_\perp, \boldsymbol{r}_\perp, x_\mathbb{P}).
\] 
Since \(|t|\) and \(\mathbf{s}_\perp\) are Fourier conjugates, the spatial distribution \(\widetilde{\mathcal{A}}(\mathbf{s}_\perp)\) provides the structural basis that governs the characteristics of \(J/\psi\) production in momentum space. To explicitly characterize the spatial extent of the interaction region, we compute the radial distribution \(|\langle \widetilde{\mathcal{A}}(b)\rangle_\Omega|\) as a function of the transverse distance \(|\mathbf{s}_\perp|\). The results shown in Fig.~\ref{fig:thickness} (b) reveal a clear transition: as the neutron-skin thickness increases, the effective nuclear color density becomes more diluted in the central region while developing a more extended tail at large radii. Comparing this with the nuclear thickness function, we can find the two figures show a similar behavior which means that the color density profile is caused by nucleon distribution in different neutron skin thickness cases. This different neutron skin thickness behavior in coordinate space is related to the \(d\sigma/dt\) in momentum space.

\begin{figure}[t]
\centering
\includegraphics[width=0.8\linewidth]{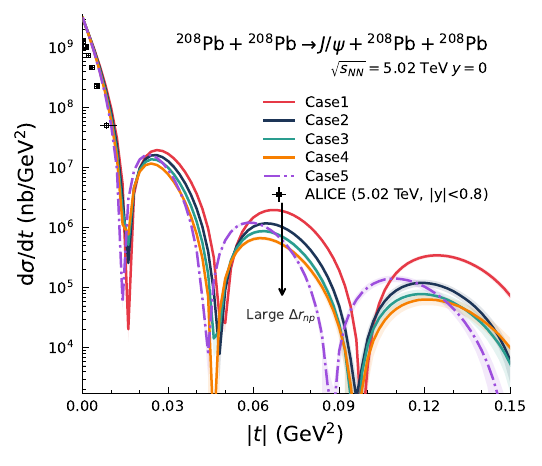}
\vspace{-0.5cm}
\caption{The coherent cross section as a function of $|t|$ for $^{208}\mathrm{Pb}+{}^{208}\mathrm{Pb}\to J/\psi+{}^{208}\mathrm{Pb}+^{208}\text{Pb}$ collisions at $\sqrtsnn=5.02\;\text{TeV}$ and $y=0$. The bands represent statistical errors. The experimental data are taken from Ref.~\cite{ALICE:2021tyx}.}
\label{fig:fig4}
\end{figure}

Fig.~\ref{fig:fig4} illustrates the impact of the neutron-skin thickness on the coherent cross section in $\text{Pb}+\text{Pb}\rightarrow J/\psi+\text{Pb}+\text{Pb}$ scattering. The results show a strong sensitivity to the neutron-skin thickness, particularly in the high-$|t|$ region ($|t|>0.05~\mathrm{GeV}^2$), where a clear suppression emerges as the neutron skin becomes thicker. For example, at $|t|=0.08~\mathrm{GeV}^2$, the baseline scenario (Case~2) yields a cross section approximately 2.14 times larger than the thick neutron-skin case (Case~4) and about 1.60 times larger than the moderate neutron-skin case (Case~3), whereas the thin neutron-skin scenario (Case~1) exceeds the baseline by a factor of 2.36. It should be noted that the proximity of the cross sections for Cases 3 and 4 in large $|t|$ region arises mainly from the shift of diffractive minima and statistical fluctuations. Our results exhibit a qualitative trend consistent with the ALICE experimental data~\cite{ALICE:2021tyx}. The discrepancy in absolute magnitude is attributed to a scaling factor, which stems from inherent theoretical uncertainties, including higher-order corrections and parameterizations of the vector meson wave function within the model. Furthermore, we evaluate Case~5, which shares the same neutron skin thickness as Case~4 but incorporates only the bulk contribution. As demonstrated in the results, isolating the bulk contribution in Case~5 leads to a shift of the diffractive minima towards lower $|t|$ compared to the scenario without a neutron skin, indicating that the coherent cross section maintains a sensitivity to the bulk effect. Furthermore, a comparison of Case 5 and Case 4 reveals that while their coherent cross sections exhibit similar behavior in the small-$|t|$ region, they can be clearly distinguished at large $|t|$ due to a visible shift in the diffractive minima. Physically, a thicker neutron skin pushes the neutron distribution to larger radii, smoothing the nuclear edge and diluting the effective color density near the surface. In momentum space, this smoother and more extended spatial profile suppresses the high-frequency components of the nuclear form factor, leading to a steeper fall-off of the coherent cross section at large $|t|$. However, we note that extracting the pure coherent signal at such $|t|>0.05\;\text{GeV}^2$ region is experimentally challenging as discussed in \cite{Chang:2021jnu}. This practical limitation naturally motivates our subsequent focus on the ratio of incoherent cross section over coherent cross section and specifically optimized $t$-integrated observables.

\begin{figure}[t]  
\centering
\includegraphics[width=0.8\linewidth]{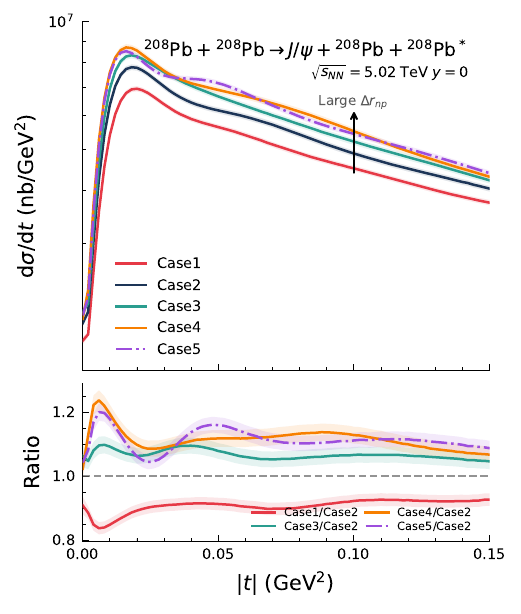}
\vspace{-0.5cm}
\caption{Upper: The incoherent cross section as a function of $|t|$ for
$^{208}\mathrm{Pb}+{}^{208}\mathrm{Pb}\to J/\psi+{}^{208}\mathrm{Pb}+{}^{208}\mathrm{Pb}^{*}$ collisions at $\sqrtsnn=5.02\;\text{TeV}$ and $y=0$.
Lower: The ratio of the incoherent cross sections for different neutron-skin thicknesses relative to Case~2, which corresponds to the scenario without neutron skin as a baseline. The bands represent statistical errors.}
\label{fig:fig5}
\end{figure}

\begin{figure}[t]
\centering
\includegraphics[width=0.8\linewidth]{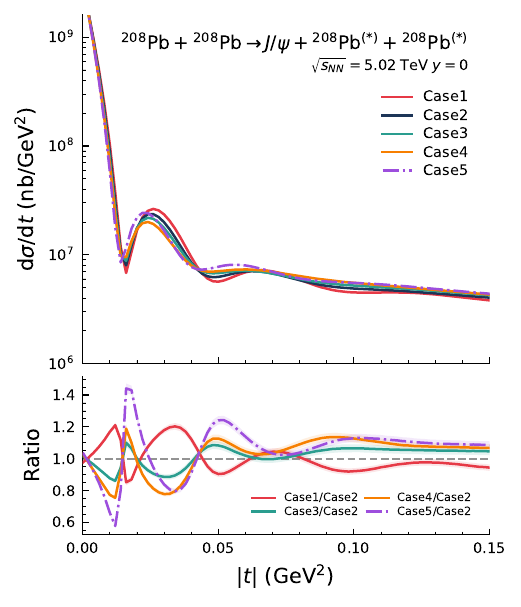}
\vspace{-0.5cm}
\caption{Upper: The total cross section as a function of $|t|$ for
$^{208}\mathrm{Pb}+{}^{208}\mathrm{Pb}\to J/\psi+{}^{208}\mathrm{Pb}^{(*)}+{}^{208}\mathrm{Pb}^{(*)}$ collisions at $\sqrtsnn=5.02\;\text{TeV}$ and $y=0$.
Lower: The ratio of the total cross sections for different neutron-skin thicknesses relative to Case~2, which corresponds to the scenario without neutron skin as a baseline. The bands represent statistical errors.}
\label{fig:fig6}
\end{figure}

Figure~\ref{fig:fig5} presents the incoherent cross section and its ratio relative to Case~2. In incoherent scattering, the sensitivity of the cross section to the neutron-skin thickness is evident over a wide range of $|t|$. As shown in the bottom panel of Fig.~\ref{fig:fig5}, the ratio demonstrates that increasing the neutron-skin thickness enhances the incoherent yield. Specifically, at $|t| = 0.006\;\mathrm{GeV}^2$, the scenario with a reduced neutron-skin thickness (Case~1) exhibits a suppression of the cross section, with a ratio of 0.83. In contrast, a moderate neutron-skin thickness (Case~3) leads to a 9.6\% increase, while a large neutron-skin thickness (Case~4) results in a 23.9\% enhancement relative to Case~2. Similarly, in Case~5, where only the bulk contribution is considered, the incoherent cross section is also enhanced relative to the no-skin baseline, implying that the bulk dynamics introduce additional fluctuations. This behavior indicates that a thicker neutron skin amplifies nuclear color-configuration fluctuations, particularly in the nuclear periphery, thereby increasing the variance of the scattering amplitude in Eq.~\eqref{eq.incoherent}.

Figure~\ref{fig:fig6} depicts the total differential cross section and its ratio relative to Case~2. At low $|t|$, the sensitivity of the total cross section closely follows that of the coherent contribution, indicating that coherent scattering dominates the total yield in this region. As $|t|$ increases, the coherent component falls off rapidly due to the suppression of high-frequency Fourier components, and incoherent scattering becomes the primary contribution. Consequently, the sensitivity of the total cross section at larger $|t|$ is governed by the incoherent channel.

\begin{figure}[htb]
\centering
\includegraphics[width=1\linewidth]{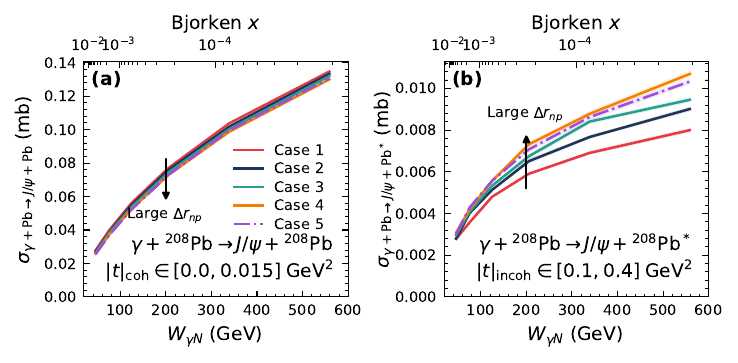}
\vspace{-0.8cm}
\caption{The integrated cross sections for (a) coherent (\(\gamma+\mathrm{Pb}\to J/\psi+\mathrm{Pb}\)) and (b) incoherent (\(\gamma+\mathrm{Pb}\to J/\psi+\mathrm{Pb}^*\)) photoproduction are shown as functions of the photon–nucleon center-of-mass energy \(W_{\gamma N}\) and Bjorken-\(x\). The \(|t|\) integration ranges are \([0,0.015]\;\mathrm{GeV}^2\) for the coherent cross section and \([0.1,0.4]\;\mathrm{GeV}^2\) for the incoherent cross section. The bands represent statistical errors.}
\label{fig:fig_Wgammap}
\end{figure}

\begin{figure}[htb]
\centering
\includegraphics[width=1\linewidth]{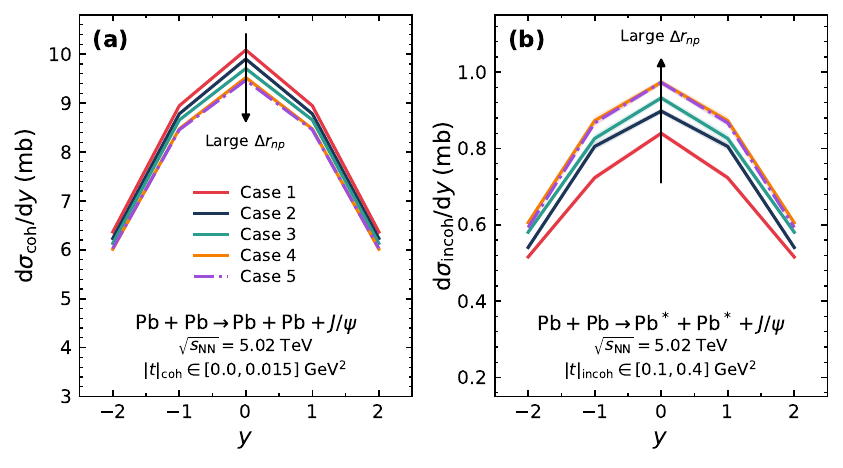}
\vspace{-0.8cm}
\caption{Rapidity distributions $d\sigma/dy$ of $J/\psi$ photoproduction in Pb-Pb UPCs at $\sqrt{s_\text{NN}} = 5.02$ TeV. (a) Coherent cross section integrated over $|t| \in [0.0, 0.015]\;\text{GeV}^2$. (b) Incoherent cross section integrated over $|t| \in [0.1, 0.4]\;\text{GeV}^2$. The bands represent statistical errors.}
\label{fig:fig_y}
\end{figure}

\begin{figure}[htb]
\centering
\includegraphics[width=0.85\linewidth]{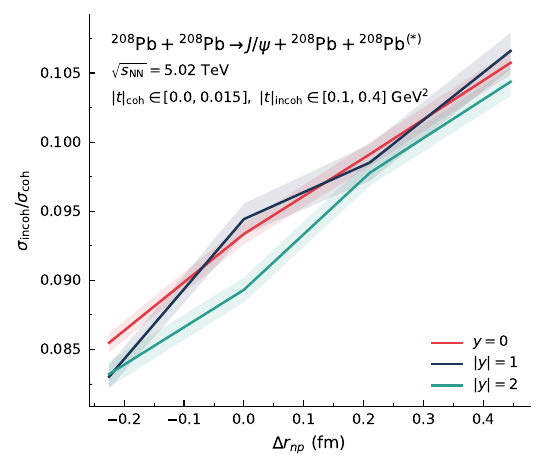}
\vspace{-0.5cm}
\caption{The ratio of the incoherent to coherent integrated cross sections as a function of the neutron-skin thickness $\Delta r_{np}$. The \(|t|\) integration ranges are \([0,0.015]\;\mathrm{GeV}^2\) for the coherent cross section and \([0.1,0.4]\;\mathrm{GeV}^2\) for the incoherent cross section. The results are shown for three different rapidities ($y=0$, $|y|=1$, and $|y|=2$).}
\label{fig:fig_ratio}
\end{figure}

Figure~\ref{fig:fig_Wgammap} presents the \(\gamma+\mathrm{Pb}\) cross section \(\sigma_{\gamma+\mathrm{Pb}\to J/\psi+\mathrm{Pb}}\) as functions of the photon–nucleon center-of-mass energy \(W_{\gamma N}\) and Bjorken-\(x\). In this calculation, we integrate over \(|t|\in[0.0,0.015]\;\mathrm{GeV}^2\) for the coherent cross section shown in panel (a), and over \(|t|\in[0.1,0.4]\;\mathrm{GeV}^2\) for the incoherent cross section shown in panel (b). Clearly, increasing the neutron-skin thickness reduces the coherent cross section and enhances the incoherent cross section.

Fig.~\ref{fig:fig_y} presents the total \(J/\psi\) production cross section as a function of rapidity, \(\mathrm{d}\sigma/\mathrm{d}y\). The results are obtained by integrating the differential cross sections over \(|t|\in[0.0,0.015]\;\mathrm{GeV}^2\) for coherent scattering in panel (a) and \(|t|\in[0.1,0.4]\;\mathrm{GeV}^2\) for incoherent scattering in panel (b). Similar to Fig.~\ref{fig:fig_Wgammap}, we find that as the neutron-skin thickness increases, the coherent cross section is reduced while the incoherent cross section is enhanced. In the present \(|t|\) region, Case~5 is almost the same as Case~4, indicating that the bulk contribution is difficult to distinguish from the surface contribution using the coherent cross section in the low-\(|t|\) region and the incoherent cross section over the whole range. In the very large-\(|t|\) region, however, the bulk contribution can be distinguished from the surface contribution by the coherent cross section, as discussed in Fig.~\ref{fig:fig4}.

Figure~\ref{fig:fig_ratio} shows the ratio of the incoherent to coherent integrated cross sections, \(\sigma_\text{Incoh}/\sigma_\text{Coh}\), as a function of the neutron-skin thickness. The coherent cross section is integrated over \(|t|\in[0.0,0.015]\;\mathrm{GeV}^2\) and the incoherent cross section over \(|t|\in[0.1,0.4]\;\mathrm{GeV}^2\). As discussed with Fig.~\ref{fig:fig4}, measuring \(\mathrm{d}\sigma/\mathrm{d}t\) at $|t|>0.05\;\text{GeV}^2$ is experimentally challenging; in contrast, \(\sigma_\text{Incoh}/\sigma_\text{Coh}\) uses a small-\(|t|\) region for the coherent part and a relatively large-\(|t|\) region for the incoherent part, both of which are accessible experimentally. This ratio largely suppresses uncertainties associated with the wave function modeling and provides a more sensitive probe of the neutron-skin thickness, since an increased neutron skin simultaneously suppresses the coherent cross section and enhances the incoherent cross section at larger \(|t|\). Furthermore, evaluating this ratio at different rapidities (\(y=0\), \(|y|=1\), and \(|y|=2\)) shows only small differences.

\section{Summary}\label{sec:summary}
In summary, we have investigated the sensitivity of coherent, incoherent, and total $J/\psi$ photoproduction cross sections in ultraperipheral Pb+Pb collisions to the neutron-skin thickness of the lead nucleus. Our results demonstrate that the neutron skin leaves a pronounced imprint on the momentum-transfer dependence of the scattering process. 

In particular, the coherent cross section exhibits strong suppression at large $|t|$ as the neutron-skin thickness increases, reflecting the smoothing of the color density profile and the corresponding suppression of high-frequency components in the nuclear form factor. In contrast, the incoherent cross section shows an overall enhancement with increasing neutron-skin thickness over a wide $|t|$ range. This enhancement can be attributed to amplified nuclear color-configuration fluctuations, especially in the nuclear periphery, which increase the variance of the scattering amplitude. As a consequence, the total cross section transitions from being dominated by coherent scattering at low $|t|$ to incoherent scattering at larger $|t|$, with the corresponding sensitivity to the neutron-skin thickness evolving accordingly. 
Finally, we show that the ratio of incoherent to coherent integrated cross sections, evaluated in experimentally measurable \(|t|\) windows—integrating the coherent cross section over \(|t|\in[0.0,0.015]~\mathrm{GeV}^2\) and the incoherent cross section over \(|t|\in[0.1,0.4]~\mathrm{GeV}^2\)—provides a robust and more sensitive probe of the neutron-skin thickness. This ratio reduces common theoretical uncertainties, has only minor rapidity dependence, and amplifies the opposite neutron-skin trends of the two channels.

Although measuring the coherent cross section at very large \(|t|\) is experimentally challenging, it could provide a potential probe to distinguish bulk and surface components of the neutron skin. Such measurements, together with the proposed ratios, would offer a valuable tomographic tool to constrain the neutron-skin thickness and the underlying spatial distribution of gluons, providing a theoretical baseline for future measurements at the LHC and future EIC.

\begin{acknowledgments}
We thank Wenbin Zhao, Chun Shen, Bjoern Schenke, Jie Zhao, and Zaochen Ye for fruitful discussions. We also thank Chen Zhong for helping with the computation server. This work is supported by the National Key Research and Development Program of China under Contracts Nos. 2024YFA1612600 and 2022YFA1604900, the National Natural Science Foundation of China under Contracts Nos. 12025501 and 12547102, Shanghai Pujiang Talents Program under Contract No. 24PJA009, the China Postdoctoral Science Foundation under Grant No. 2024M750489, and the Natural Science Foundation of Shanghai under Grant No. 23JC1400200.
\end{acknowledgments}


\begin{thebibliography}{99}

\bibitem{Steiner:2004fi}
Steiner A~W, Prakash M, Lattimer J~M and Ellis P~J 2005 \href{https://doi.org/10.1016/j.physrep.2005.02.004}{{\em Phys. Rept.\/} {\bf 411} 325--375}

\bibitem{Lattimer:2006xb}
Lattimer J~M and Prakash M 2007 \href{https://doi.org/10.1016/j.physrep.2007.02.003}{{\em Phys. Rept.\/} {\bf 442} 109--165}

\bibitem{Li:2008gp}
Li B~A, Chen L~W and Ko C~M 2008 \href{https://doi.org/10.1016/j.physrep.2008.04.005}{{\em Phys. Rept.\/} {\bf 464} 113--281}

\bibitem{Zhang:2017ncy}
Zhang N~B, Cai B~J, Li B~A, Newton W~G and Xu J 2017 \href{https://doi.org/10.1007/s41365-017-0336-2}{{\em Nucl. Sci. Tech.\/} {\bf 28} 181}

\bibitem{Chen:2010qx}
Chen L~W, Ko C~M, Li B~A and Xu J 2010 \href{https://doi.org/10.1103/PhysRevC.82.024321}{{\em Phys. Rev. C\/} {\bf 82} 024321}

\bibitem{An_NST}
An R, Sun S, Cao L~G and Zhang F~S 2024 \href{https://doi.org/10.1007/s41365-024-01551-w}{{\em Nucl. Sci. Tech.\/} {\bf 35} 182}

\bibitem{Hu:2021trw}
Hu B {\em et~al.\/} 2022 \href{https://doi.org/10.1038/s41567-023-02324-9}{{\em Nature Phys.\/} {\bf 18} 1196--1200}

\bibitem{PREX:2021umo}
Adhikari D {\em et~al.\/} (PREX Collaboration) 2021 \href{https://doi.org/10.1103/PhysRevLett.126.172502}{{\em Phys. Rev. Lett.\/} {\bf 126}(17) 172502}

\bibitem{Kumar:2020ejz}
Kumar K~S (PREX, CREX) 2020 \href{https://doi.org/10.1016/j.aop.2019.168012}{{\em Annals Phys.\/} {\bf 412} 168012}

\bibitem{Ma:2022dbh}
Ma Y~G and Zhang S 2022 \href{https://doi.org/10.1007/978-981-15-8818-1_5-1}{{\em {Influence of Nuclear Structure in Relativistic Heavy-Ion Collisions}\/} pp 1--30}

\bibitem{STAR:2024wgy}
Abdulhamid M~I {\em et~al.\/} (STAR) 2024 \href{https://doi.org/10.1038/s41586-024-08097-2}{{\em Nature\/} {\bf 635} 67--72}

\bibitem{STAR:2025elk}
Aboona B~E {\em et~al.\/} (STAR) 2025 \href{https://doi.org/10.1088/1361-6633/ae0fc3}{{\em Rept. Prog. Phys.\/} {\bf 88} 108601}

\bibitem{Zhang:2021kxj}
Zhang C and Jia J 2022 \href{https://doi.org/10.1103/PhysRevLett.128.022301}{{\em Phys. Rev. Lett.\/} {\bf 128} 022301}

\bibitem{Jia:2021oyt}
Jia J and Zhang C 2023 \href{https://doi.org/10.1103/PhysRevC.107.L021901}{{\em Phys. Rev. C\/} {\bf 107} L021901}

\bibitem{Chen:2024aom}
Chen J {\em et~al.\/} 2024 \href{https://doi.org/10.1007/s41365-024-01591-2}{{\em Nucl. Sci. Tech.\/} {\bf 35} 214}

\bibitem{Chen:2026gka}
Chen J {\em et~al.\/} 2026 \href{https://arxiv.org/abs/2601.12977}{arXiv: 2601.12977} [nucl-ex]

\bibitem{Xi_NST}
Xi B~S, Chen J~H, Ma L, Ma Y~G and Wang T~T 2025 \href{https://doi.org/10.1007/s41365-025-01826-w}{{\em Nucl. Sci. Tech.\/} {\bf 36} 228}

\bibitem{Giacalone:2024bud}
Giacalone G 2024 \href{https://doi.org/10.1007/s41365-024-01582-3}{{\em Nucl. Sci. Tech.\/} {\bf 35} 218}

\bibitem{SB_NST}
Schenke B 2024 \href{https://doi.org/10.1007/s41365-024-01509-y}{{\em Nucl. Sci. Tech.\/} {\bf 35} 115}

\bibitem{Fang:2010zzc}
Fang D~Q, Ma Y~G, Cai X~Z, Tian W~D and Wang H~W 2010 \href{https://doi.org/10.1103/PhysRevC.81.047603}{{\em Phys. Rev. C\/} {\bf 81} 047603}

\bibitem{Li:2016gjs}
Li X~F, Fang D~Q and Ma Y~G 2016 \href{https://doi.org/10.1007/s41365-016-0064-z}{{\em Nucl. Sci. Tech.\/} {\bf 27} 71}

\bibitem{Ding:2024xxu}
Ding M~Q, Fang D~Q and Ma Y~G 2024 \href{https://doi.org/10.1007/s41365-024-01584-1}{{\em Nucl. Sci. Tech.\/} {\bf 35} 211}

\bibitem{MaCW_NST}
Ma C~W, Duan Y~J, Guo Y~F, Qiao C~Y, Wang Y~T, Pu J, Cheng K~X and Wei H~L 2024 \href{https://doi.org/10.1007/s41365-024-01455-9}{{\em Nucl. Sci. Tech.\/} {\bf 35} 99}

\bibitem{Yan_NST}
Yan T~Z and Li S 2024 \href{https://doi.org/10.1007/s41365-024-01425-1}{{\em Nucl. Sci. Tech.\/} {\bf 35} 65}

\bibitem{Li:2024fsx}
Li T~Z, Liu L~M, Xu J and Ren Z~Z 2024 \href{https://doi.org/10.1103/PhysRevC.110.054613}{{\em Phys. Rev. C\/} {\bf 110} 054613}

\bibitem{Jia:2022qgl}
Jia J, Giacalone G and Zhang C 2023 \href{https://doi.org/10.1103/PhysRevLett.131.022301}{{\em Phys. Rev. Lett.\/} {\bf 131} 022301}

\bibitem{Giacalone:2023cet}
Giacalone G, Nijs G and van~der Schee W 2023 \href{https://doi.org/10.1103/PhysRevLett.131.202302}{{\em Phys. Rev. Lett.\/} {\bf 131} 202302}

\bibitem{Li:2019kkh}
Li H, Xu H~j, Zhou Y, Wang X, Zhao J, Chen L~W and Wang F 2020 \href{https://doi.org/10.1103/PhysRevLett.125.222301}{{\em Phys. Rev. Lett.\/} {\bf 125} 222301}

\bibitem{Liu:2023pav}
Liu Q, Zhao S, Xu H~J and Song H 2024 \href{https://doi.org/10.1103/PhysRevC.109.034912}{{\em Phys. Rev. C\/} {\bf 109} 034912}

\bibitem{Liu:2022xlm}
Liu L~M, Zhang C~J, Xu J, Jia J and Peng G~X 2022 \href{https://doi.org/10.1103/PhysRevC.106.034913}{{\em Phys. Rev. C\/} {\bf 106} 034913}

\bibitem{Liu:2022kvz}
Liu L~M, Zhang C~J, Zhou J, Xu J, Jia J and Peng G~X 2022 \href{https://doi.org/10.1016/j.physletb.2022.137441}{{\em Phys. Lett. B\/} {\bf 834} 137441}

\bibitem{Liu:2023qeq}
Liu L~M, Xu J and Peng G~X 2023 \href{https://doi.org/10.1016/j.physletb.2023.137701}{{\em Phys. Lett. B\/} {\bf 838} 137701}

\bibitem{Jia:2022ozr}
Jia J {\em et~al.\/} 2024 \href{https://doi.org/10.1007/s41365-024-01589-w}{{\em Nucl. Sci. Tech.\/} {\bf 35} 220}

\bibitem{Sorensen:2023zkk}
Sorensen A {\em et~al.\/} 2024 \href{https://doi.org/10.1016/j.ppnp.2023.104080}{{\em Prog. Part. Nucl. Phys.\/} {\bf 134} 104080}

\bibitem{STAR:2021wwq}
Abdallah M {\em et~al.\/} (STAR) 2022 \href{https://doi.org/10.1103/PhysRevLett.128.122303}{{\em Phys. Rev. Lett.\/} {\bf 128} 122303}

\bibitem{STAR:2022wfe}
Abdallah M {\em et~al.\/} (STAR) 2023 \href{https://doi.org/10.1126/sciadv.abq3903}{{\em Sci. Adv.\/} {\bf 9} eabq3903}

\bibitem{Abelev_2014}
Abelev B~B {\em et~al.\/} (ALICE) 2014 \href{https://doi.org/10.1103/PhysRevLett.113.232504}{{\em Phys. Rev. Lett.\/} {\bf 113} 232504}

\bibitem{CMS:2025oxg}
Chekhovsky V {\em et~al.\/} (CMS) 2025 \href{https://doi.org/10.1103/w9kp-f8xr}{{\em Phys. Rev. Lett.\/} {\bf 135} 112301}

\bibitem{ATLAS:2025aav}
Aad G {\em et~al.\/} (ATLAS) 2025  \href{https://arxiv.org/abs/2509.04135}{arXiv: 2509.04135} [nucl-ex]

\bibitem{ATLAS:2025nac}
Aad G {\em et~al.\/} (ATLAS) 2025  \href{https://arxiv.org/abs/2504.07795}{arXiv: 2504.07795} [nucl-ex]

\bibitem{ALICE:2024whv}
Acharya S {\em et~al.\/} (ALICE) 2025 \href{https://doi.org/10.1016/j.physletb.2025.139952}{{\em Phys. Lett. B\/} {\bf 871} 139952}

\bibitem{CMS:2023pzm}
CMS Collaboration 2023 \href{https://doi.org/10.1016/j.physletb.2023.138131}{{\em Phys. Lett. B\/} {\bf 845} 138131}

\bibitem{ALICE:2023jgu}
Acharya S {\em et~al.\/} (ALICE) 2023 \href{https://doi.org/10.1007/JHEP10(2023)119}{{\em JHEP\/} {\bf 10} 119}

\bibitem{Brandenburg:2025one}
Brandenburg J~D, Klein S~R, Xu Z, Yang S, Zha W and Zhou J 2025 \href{https://doi.org/10.1016/j.ppnp.2025.104174}{{\em Prog. Part. Nucl. Phys.\/} {\bf 143} 104174}

\bibitem{Shen_Res}
Shen D~Y, Chen J~H, Huang X~G, Ma Y~G, Tang A~H and Wang G 2025 \href{https://doi.org/10.34133/research.0726}{{\em Research\/} {\bf 8} 0726}

\bibitem{Jia:2025oak}
Jia Y, Sang W~L, Xiong X, Zhou J and Zhou Y~J 2025 \href{https://arxiv.org/abs/2512.23306}{arXiv: 2512.23306} [hep-ph]

\bibitem{Baur2008}
Baur G 2008 \href{https://doi.org/10.1016/j.nuclphysbps.2008.07.017}{{\em Nucl. Phys. B Proc. Suppl.\/} {\bf 179-180} 129--133}

\bibitem{Baur:1998ay}
Baur G, Hencken K and Trautmann D 1998 \href{https://doi.org/10.1088/0954-3899/24/9/003}{{\em J. Phys. G\/} {\bf 24} 1657--1692}

\bibitem{starlight_Klein_2017}
Klein S~R, Nystrand J, Seger J, Gorbunov Y and Butterworth J 2017 \href{https://doi.org/10.1016/j.cpc.2016.10.016}{{\em Computer Physics Communications\/} {\bf 212} 258--268}

\bibitem{Klein:2019qfb}
Klein S~R and M{\"a}ntysaari H 2019 \href{https://doi.org/10.1038/s42254-019-0107-6}{{\em Nature Rev. Phys.\/} {\bf 1} 662--674}

\bibitem{Klein:2019avl}
Klein S~R and Xie Y~P 2019 \href{https://doi.org/10.1103/PhysRevC.100.024620}{{\em Phys. Rev. C\/} {\bf 100} 024620}

\bibitem{Lomnitz:2018juf}
Lomnitz M and Klein S 2019 \href{https://doi.org/10.1103/PhysRevC.99.015203}{{\em Phys. Rev. C\/} {\bf 99} 015203}

\bibitem{chun23}
M{\"a}ntysaari H, Schenke B, Shen C and Zhao W 2023 \href{https://doi.org/10.1103/PhysRevLett.131.062301}{{\em Phys. Rev. Lett.\/} {\bf 131} 062301}

\bibitem{mantysaari2023}
M{\"a}ntysaari H, Salazar F, Schenke B, Shen C and Zhao W 2024 \href{https://doi.org/10.1103/PhysRevC.109.024908}{{\em Phys. Rev. C\/} {\bf 109} 024908}

\bibitem{Zhao:2022dac}
Zhao J, Chen J~H, Huang X~G and Ma Y~G 2024 \href{https://doi.org/10.1007/s41365-024-01374-9}{{\em Nucl. Sci. Tech.\/} {\bf 35} 20}

\bibitem{Wang:2024xeo}
Wang J, Chen B and Liu Y 2024 \href{https://doi.org/10.1088/0256-307X/41/10/102501}{{\em Chin. Phys. Lett.\/} {\bf 41} 102501}

\bibitem{Accardi:2012qut}
Accardi A {\em et~al.\/} 2016 \href{https://doi.org/10.1140/epja/i2016-16268-9}{{\em Eur. Phys. J. A\/} {\bf 52} 268}

\bibitem{Anderle:2021wcy}
Anderle D~P {\em et~al.\/} 2021 \href{https://doi.org/10.1007/s11467-021-1062-0}{{\em Front. Phys. (Beijing)\/} {\bf 16} 64701}

\bibitem{Chu:2024fpo}
Chu Z, Chen J, Wang X~P and Xing H 2025 \href{https://doi.org/10.1103/PhysRevD.111.L011501}{{\em Phys. Rev. D\/} {\bf 111} L011501}

\bibitem{Wu_NST}
Wu X, Li X~B, Tang Z~B, Wang K~Y and Zha W~M 2025 \href{https://doi.org/10.1007/s41365-025-01704-5}{{\em Nucl. Sci. Tech.\/} {\bf 36} 157}

\bibitem{AbdulKhalek:2021gbh}
Abdul~Khalek R {\em et~al.\/} 2022 \href{https://doi.org/10.1016/j.nuclphysa.2022.122447}{{\em Nucl. Phys. A\/} {\bf 1026} 122447}

\bibitem{Zhang:2025raf}
Zhang S~L, Wang E, Wang X~N and Xing H 2025 \href{https://arxiv.org/abs/2506.10694}{arXiv: 2506.10694} [hep-ph]

\bibitem{Boer:2011fh}
Boer D {\em et~al.\/} 2011 \href{https://arxiv.org/abs/1108.1713}{arXiv: 1108.1713} [nucl-th]

\bibitem{Gimeno-Estivill:2024gbu}
Gimeno-Estivill P, Lappi T and M{\"a}ntysaari H 2024 \href{https://doi.org/10.1103/PhysRevD.110.094035}{{\em Phys. Rev. D\/} {\bf 110} 094035}

\bibitem{Mantysaari:2022sux}
M{\"a}ntysaari H, Salazar F and Schenke B 2022 \href{https://doi.org/10.1103/PhysRevD.106.074019}{{\em Phys. Rev. D\/} {\bf 106} 074019}

\bibitem{Bartels:2003yj}
Bartels J, Golec-Biernat K and Peters K 2003 \href{https://arxiv.org/abs/hep-ph/0301192}{arXiv: hep-ph/0301192} [hep-ph]

\bibitem{Watt:2008vq}
Kowalski H, Motyka L and Watt G 2006 \href{https://doi.org/10.1103/PhysRevD.74.074016}{{\em Phys. Rev. D\/} {\bf 74} 074016}

\bibitem{Bertulani:2005ru}
Bertulani C~A, Klein S~R and Nystrand J 2005 \href{https://doi.org/10.1146/annurev.nucl.55.090704.151526}{{\em Ann. Rev. Nucl. Part. Sci.\/} {\bf 55} 271--310}

\bibitem{PhysRev.120.1857}
Good M~L and Walker W~D 1960 \href{https://doi.org/10.1103/PhysRev.120.1857}{{\em Phys. Rev.\/} {\bf 120}(5) 1857--1860}

\bibitem{Klein:2023zlf}
Klein S~R 2023 \href{https://doi.org/10.1103/PhysRevC.107.055203}{{\em Phys. Rev. C\/} {\bf 107} 055203}

\bibitem{Mantysaari:2020axf}
M{\"a}ntysaari H 2020 \href{https://doi.org/10.1088/1361-6633/aba347}{{\em Rept. Prog. Phys.\/} {\bf 83} 082201}

\bibitem{PhysRevD.18.1696}
Miettinen H~I and Pumplin J 1978 \href{https://doi.org/10.1103/PhysRevD.18.1696}{{\em Phys. Rev. D\/} {\bf 18}(5) 1696--1708}

\bibitem{Iancu_2004}
Iancu E and Venugopalan R 2003 \href{https://doi.org/10.1142/9789812795533_0005}{{\em {The Color glass condensate and high-energy scattering in QCD}\/} pp 249--3363}

\bibitem{Gelis_2010}
Gelis F, Iancu E, Jalilian-Marian J and Venugopalan R 2010 \href{https://doi.org/10.1146/annurev.nucl.010909.083629}{{\em Ann. Rev. Nucl. Part. Sci.\/} {\bf 60} 463--489}

\bibitem{McLerran_2011}
McLerran L 2011 \href{https://doi.org/10.1143/PTPS.187.17}{{\em Prog. Theor. Phys. Suppl.\/} {\bf 187} 17--30}

\bibitem{Ji:1996nm}
Ji X 1997 \href{https://doi.org/10.1103/PhysRevD.55.7114}{{\em Phys. Rev. D\/} {\bf 55}(11) 7114--7125}

\bibitem{Hatta:2017cte}
Hatta Y, Xiao B~W and Yuan F 2017 \href{https://doi.org/10.1103/PhysRevD.95.114026}{{\em Phys. Rev. D\/} {\bf 95}(11) 114026}

\bibitem{Kowalski_2006}
Kowalski H, Motyka L and Watt G 2006 \href{https://doi.org/10.1103/PhysRevD.74.074016}{{\em Phys. Rev. D\/} {\bf 74}(7) 074016}

\bibitem{Kovchegov22}
Kovchegov Y~V and Levin E 2022 \href{https://doi.org/10.1017/9781009291446}{{\em Quantum Chromodynamics at High Energy\/} ({\em Cambridge Monographs on Particle Physics, Nuclear Physics and Cosmology\/} vol~33) (Cambridge University Press)}

\bibitem{Mantysaari:2017dwh}
M{\"a}ntysaari H and Schenke B 2017 \href{https://doi.org/10.1016/j.physletb.2017.07.063}{{\em Phys. Lett. B\/} {\bf 772} 832--838}

\bibitem{Mantysaari:2025ltq}
M{\"a}ntysaari H, Roch H, Salazar F, Schenke B, Shen C and Zhao W 2026 \href{https://doi.org/10.1103/pcmz-dyz1}{{\em Phys. Rev. D\/} {\bf 113} 014038}

\bibitem{Warda:2010qa}
Warda M, Vinas X, Roca-Maza X and Centelles M 2010 \href{https://doi.org/10.1103/PhysRevC.81.054309}{{\em Phys. Rev. C\/} {\bf 81} 054309}

\bibitem{ipglasma_code}
IP-Glasma 2024 \url{https://github.com/chunshen1987/IPGlasmaFramework}

\bibitem{subnucleondiffraction_code}
M{\"a}ntysaari H 2024 \url{https://github.com/hejajama/subnucleondiffraction}

\bibitem{Schenke:2012wb}
Schenke B, Tribedy P and Venugopalan R 2012 \href{https://doi.org/10.1103/PhysRevLett.108.252301}{{\em Phys. Rev. Lett.\/} {\bf 108}(25) 252301}

\bibitem{Schenke:2012hg}
Schenke B, Tribedy P and Venugopalan R 2012 \href{https://doi.org/10.1103/PhysRevC.86.034908}{{\em Phys. Rev. C\/} {\bf 86}(3) 034908}

\bibitem{ALICE:2021tyx}
Acharya S {\em et~al.\/} (ALICE) 2021 \href{https://doi.org/10.1016/j.physletb.2021.136280}{{\em Phys. Lett. B\/} {\bf 817} 136280}

\bibitem{Chang:2021jnu}
Chang W, Aschenauer E~C, Baker M~D, Jentsch A, Lee J~H, Tu Z, Yin Z and Zheng L 2021 \href{https://doi.org/10.1103/PhysRevD.104.114030}{{\em Phys. Rev. D\/} {\bf 104} 114030}

\end{thebibliography}
\end{document}